\documentclass[12pt]{iopart}
\usepackage{graphicx}% Include figure files

\begin{document}

\title{Heavy Quarkonia and Quark Drip Lines in Quark-Gluon Plasma}

\author{Cheuk-Yin Wong}
\address
{Physics Division, Oak Ridge National Laboratory, Oak Ridge, TN
37831}
\address
{\& Department of Physics, University of Tennessee, Knoxville, TN 
37996}
\vspace*{0.2cm}
\address{E-mail: wongc@ornl.gov} 

\begin{abstract}
Using the potential model and thermodynamical quantities obtained in
lattice gauge calculations, we determine the spontaneous dissociation
temperatures of color-singlet quarkonia and the `quark drip lines'
which separate the region of bound $Q\bar Q$ states from the unbound
region.  The dissociation temperatures of $J/\psi$ and $\chi_b$ in
quenched QCD are found to be 1.62$T_c$ and $1.18T_c$ respectively, in
good agreement with spectral function analyses.  The dissociation
temperature of $J/\psi$ in full QCD with 2 flavors is found to be
1.42$T_c$.  For possible bound quarkonium states with light quarks,
the characteristics of the quark drip lines severely limit the stable
region close to the phase transition temperature.  Bound color-singlet
quarkonia with light quarks may exist very near the phase transition
temperature if their effective quark mass is of the order of 300-400
MeV and higher.
\end{abstract}
\vspace*{-0.7cm}

%\maketitle

\section{Introduction}                                                                            
The degree to which the constituents of a quark-gluon plasma can
combine to form composite entities is an important property of the
plasma.  It has significant implications on the plasma equation of
state, the probability of recombination of plasma constituents prior
to the phase transition, and the chemical yields of the observed
hadrons.  Recent spectral analyses of quarkonium correlators indicated
that $J/\psi$ may be bound up to 1.6$T_c$ where $T_c$ is the phase
transition temperature \cite{Asa03,Dat03}.  Subsequently, there has
been renewed interest in quarkonium states in quark-gluon plasma as
Zahed and Shuryak suggested that $Q \bar Q$ states with light quarks
may be bound up to a few $T_c$ \cite{Shu03}.  Quarkonium bound states
and instanton molecules in the quark-gluon plasma have been considered
by Brown, Lee, Rho, and Shuryak \cite{Bro04}.  As heavy quarkonia may
be used as diagnostic tools \cite{Mat86}, there have been many recent
investigations on the stability of heavy quarkonia in the plasma
\cite{Won04,Won05,Dig01a,Won01a,Kac02,Kac03,Kac05,Alb05,Bla05,Pet05,Dig05}.

Previously, DeTar \cite{Det85}, Hansson, Lee, and Zahed \cite{Han88},
and Simonov \cite{Sim95} observed that the range of the strong
interaction is not likely to change drastically across the phase
transition and suggested the possible existence of relatively narrow
low-lying $Q\bar Q$ states in the plasma.  On the other hand, Hatsuda
and Kunihiro \cite{Hat85} considered the persistence of soft modes in
the plasma which may manifest themselves as pion-like and sigma-like
states.  The use of the baryon-strangeness correlation and the charge
fluctuation to study the abundance of light quarkonium states in the
plasma have been suggested recently \cite{Koc05,Kar05}.

We would like to use the potential model to investigate the composite
properties of the plasma and to determine its `quark drip lines'.
Here we follow Werner and Wheeler \cite{Wer58} and use the term `drip
line' to separate the region of bound color-singlet $Q\bar Q$ states
from the unbound region of spontaneous quarkonium dissociation.  It
should be emphasized that a quarkonium can be dissociated by collision
with constituent particles to lead to the corresponding
`thermal-dissociation line' and `particle-dissociation line', which
can be interesting subjects for future investigations.

\section{The color-singlet $Q$-$\bar Q$ potential}

The most important physical quantity in the potential model is the
$Q$-$\bar Q$ potential between the quark $Q$ and the antiquark $\bar
Q$ in a color-sinbglet state. Previous works in the potential model
use the color-singlet free energy $F_1$ \cite{Dig01a,Won01a,Bla05} or
the color-singlet internal energy $U_1$ \cite{Kac02,Shu03,Alb05}
obtained in lattice gauge calculations as the color-singlet $Q$-$\bar
Q$ potential without rigorous theoretical justifications.  The
internal energy $U_1$ is significantly deeper and spatially more
extended than the free energy $F_1$. Treating the internal energy
$U_1$ as the $Q$-$\bar Q$ potential led Shuryak and Zahed to suggest
the possibility of bound color-singlet quarkonium states with light
quarks in the plasma \cite{Shu03}.  The conclusions will be quite
different if one uses the free energy $F_1$ as the $Q$-$\bar Q$
potential.

While $F_1$ or $U_1$ can both be used as the $Q$-$\bar Q$ potential at
$T=0$ (at which $F_1=U_1$), the situation is not so clear in a
thermalized quark-gluon plasma.  In Ref.\ \cite{Won04}, we show that
the equation of motion for the $Q \bar Q$ state can be obtained in two
steps.  First, one considers a lattice gauge calculation for a static
pair of color-singlet $Q$ and $\bar Q$ at a fixed separation $R$
at the temperature $T$ and obtains the free energy $F_1(R)$, the
internal energy $U_1(R)$, and their difference $TS_1({\bf
r})=U_1-F_1$.  In the second step, one considers a $Q$ and a $\bar Q$
in dynamical motion in the quark-gluon plasma.  The dynamical degrees
of freedom can be taken to be the set of quarkonium and plasma
constituent wave functions and their corresponding state occupation
numbers.  The equilibrium of such a thermalized system at a fixed
temperature and volume occurs when the grand potential is a minimum.
One explicitly writes out the grand potential, which is the sum of the
free energy and appropriate Lagrange multiplier terms, in terms of the
quarkonium and plasma constituent wave functions, the state occupation
numbers, and the mutual $Q$-$\bar Q$, $Q$-(constituent), $\bar
Q$-(constituent), and (constituent)-(constituent) interactions.  The
minimization of the grand potential with respect to a color-singlet
quarkonium wave function leads to the Schr\" odinger equation for the
quarkonium state with a color-singlet potential $U_{Q\bar Q}^{(1)}$
containing only $Q$-$\bar Q$, $Q$-(constituent), and $\bar
Q$-(constituent) interactions.  On the other hand, the internal energy
$U_1$ in the lattice gauge calculation contains not only contributions
from these $Q$-$\bar Q$, $Q$-(constituent), and $\bar Q$-(constituent)
interactions, but also an additional contribution from the
(constituent)-(constituent) interaction.  Therefore, the color-singlet
potential $U_{Q\bar Q}^{(1)}$ is determined by the internal energy
$U_1$ after subtracting out the quark-gluon plasma internal energy.
These results are further supported by the presence of a similar
relationship between the total internal energy and the $Q$-$\bar Q$
potential in the analogous case of Debye screening, for which
analytical expressions for various thermodynamical quantities can be
readily obtained and compared \cite{Won06}.

\section{The color-singlet $Q$-$\bar Q$ potential in quark-gluon plasma}

In order to subtract out the $R$-dependent internal energy of the
quark-gluon plasma from the total internal energy to obtain the
color-singlet $Q$-$\bar Q$ potential in the plasma, additional lattice
gauge calculations may be needed to evaluate the quark-gluon plasma
internal energy in the presence of a color-singlet $Q$ and $\bar Q$
pair.  It is nonetheless useful at this stage to suggest approximate
ways to evaluate this quark-gluon plasma internal energy.  The
subtraction can be carried out by noting that the quark-gluon plasma
internal energy density $\epsilon$ is related to its pressure $p$ and
entropy density $\sigma$ by the First Law of Thermodynamics,
$\epsilon=T\sigma -p$, and the quark-gluon plasma pressure $p$ is also
related to the plasma energy density $\epsilon$ by the equation of
state $p(\epsilon)$ that is presumed known by another independent
lattice gauge calculation.  Thus, by expressing $p$ as $(3p/\epsilon)
(\epsilon/3)$ with the ratio $a(T)=3p/\epsilon$ given by the known
equation of state, the plasma internal energy density $\epsilon$ is
related to the entropy density $T\sigma$ by
$\epsilon=[3/(3+a(T))]T\sigma$, and the plasma internal energy
integrated over the volume is given by $[3/(3+a(T))]TS_1$ where $TS_1$
has already been obtained as $U_1-F_1$.  The proper $Q$-$\bar Q$
potential, $U_{Q\bar Q}^{(1)}$, as determined from $U_1$ by
subtracting the plasma internal energy (which is equal to
$[3/(3+a(T))](U_1-F_1)$), is then a linear combination of $F_1$ and
$U_1$ given by \cite{Won04},
\begin{eqnarray}
\label{Uqq1}
U_{Q\bar Q}^{(1)}(R,T)= 
 \frac{3}{3+a(T)}    F_1(R,T) 
+\frac{a(T)}{3+a(T)} U_1(R,T).
\end{eqnarray}
The potential $U_{Q\bar Q}^{(1)}$ is approximately $F_1$ near $T_c$
and $3F_1/4+U_1/4$ for $T> 1.5 T_c$ \cite{Won04}.

In the spectral function analyses, the widths of many color-singlet
heavy quarkonium broaden suddenly at various temperatures
\cite{Asa03,Dat03,Pet05}.  In the most precise calculations for
$J/\psi$ in quenched QCD using up to 128 time-like lattice slices, the
spectrum has a sharp peak for $0.78T_c \le T \le 1.62T_c$ and a broad
structure with no sharp peak for $1.70T_c \le T \le 2.33T_c$
\cite{Asa03}.  The spectral peak at the bound state has the same
structure and shape at 0.78$T_c$ as it is at $1.62T_c$.  If one can
infer that $J/\psi$ is stable and bound at 0.78$T_c$, then it would be
reasonable to infer that $J/\psi$ is also bound and stable at
1.62$T_c$.  The spectral function at $1.70$$T_c$ has the same
structure and shape as the spectral function at 2.33$T_c$.  If one can
infer that $J/\psi$ is unbound at 2.33$T_c$, then it would be
reasonable to infer that $J/\psi$ become already unbound at 1.70
$T_c$.  Thus, from the shape of the spectral functions, the
temperature at which the $J/\psi$ width broadens suddenly from
1.62$T_c$ to $1.70T_c$ corresponds to the $J/\psi$ spontaneous
dissociation temperature.  Dissociation temperatures for $\chi_c$ and
$\chi_b$ in spectral analyses in quenched QCD have also been obtained
\cite{Dat03,Pet05}.  We list the heavy quarkonium spontaneous
dissociation temperatures obtained from spectral analyses in quenched
QCD in Table I.  They can be used to test the potential model of
$U_{Q\bar Q}^{(1)}$, the linear combination of $U_1$ and $F_1$
proposed in Eq.\ (\ref{Uqq1}), as well as $F_1$ and $U_1$.

\vskip 0.4cm \centerline{Table I.  Spontaneous dissociation
temperatures obtained from different potentials.}  
{\vskip 0.3cm\hskip -0.9cm
\begin{tabular}{|c|c|c|c|c|c|c|c|}
\hline
     & \multicolumn{4}{c|}{Quenched QCD} 
     & \multicolumn{3}{c|}{Full QCD (2 flavors)}       \\  \cline{2-7}
\hline
{\rm States     } & {\rm Spectral} 
                 & $U_{Q\bar Q}^{(1)}$
                 &   $F_1$  &  $U_1$ 
                 & $U_{Q\bar Q}^{(1)}$ 
                 &   $F_1$  &  $U_1$ \\

%{\rm      } & ~{\rm Spectral}~ 
%                 &~ $U_{Q\bar Q}^{(1)}(R,T)$ 
%                 &  ~ $F_1(R,T)$  & ~ $U_1(R,T)$ 
%                 &~ $U_{Q\bar Q}^{(1)}(R,T)$ 
%                 &  ~ $F_1(R,T)$  & ~ $U_1(R,T)$ \\

                 & {\rm Analysis} & 
                 &              & 
                 &              &
                 &   \\
%{\rm States} & ~{\rm Analysis}~ & {\rm Potential} 
%                 & {\rm Potential}  & {\rm Potential}
%                 & {\rm Potential}  & {\rm Potential}
%                 & {\rm Potential}  \\
\hline
$J/\psi,\eta_c$       &   $1.62$-$1.70T_c^\dagger$ 
&   $1.62\,T_c$  
&   $1.40 \,T_c$ & $2.60 \,T_c$              
&   $1.42\,T_c$  
&   $1.21 \,T_c$ & $2.22 \,T_c$              
\\ \cline{1-8}  
$\chi_c$              & below $ 1.1T_c^{\natural}$   
&   unbound
& unbound             &  $1.18 \,T_c$            
&   $1.05 \,T_c$
&   unbound           &  $1.17 \,T_c$            
\\ \cline{1-8}
%\hline
$\psi',\eta_c'$       &  
&   unbound 
&   unbound & $1.23 \,T_c$              
&   unbound  
&   unbound & $1.11 \,T_c$              
\\ \cline{1-8}  
$\Upsilon,\eta_b$     &                   
&  $ 4.1 \,T_c$    
&    $ 3.5 \,T_c$   &    $ \sim 5.0  \,T_c$                      
&  $ 3.40 \,T_c$    
&    $ 2.90 \,T_c$   &    $ 4.18 \,T_c$                           
\\ \cline{1-8}  
$\chi_b$              &  $1.15$-$1.54T_c^{\sharp}$     
&  $ 1.18 \,T_c$      
&  $ 1.10 \,T_c$      & $ 1.73 \,T_c$            
&  $ 1.22 \,T_c$      
&    $ 1.07 \,T_c$    & $ 1.61 \,T_c$            
\\ \cline{1-8}
$\Upsilon',\eta_b'$     &                   
&  $ 1.38 \,T_c$    
&    $ 1.19  \,T_c$   &    $ 2.28  \,T_c$                      
&  $ 1.18 \,T_c$    
&    $ 1.06 \,T_c$   &    $ 1.47 \,T_c$                           
\\ \cline{1-8}
                                                        \hline
\multicolumn{8}{l}
{${}^\dagger$Ref.\cite{Asa03},~~~     
 ${}^{\natural}$Ref.\cite{Dat03},~~~     
 ${}^{\sharp }$Ref.\cite{Pet05}}     
\end{tabular}
}

\vspace*{0.3cm}

\section{Quark drip-lines in quark-gluon plasma}

To evaluate the $Q$-$\bar Q$ potential, we use the free energy $F_1$
and the internal energy $U_1$ obtained by Kaczmarek $et~al.$ in both
quenched QCD \cite{Kac03} and full QCD with 2 flavors \cite{Kac05}.
In quenched QCD, $F_1$ and $U_1$ can be parametrized in terms of a
screened Coulomb potential with parameters shown in Figs. 1 and 2 of
Ref.\ \cite{Won04}.  In full QCD with 2 flavors, $F_1$ and $U_1$ can
be represented by a color-Coulomb interaction at short distances and a
completely screened, constant, potential at large distances as given
in Ref. \cite{Won05}, although other alternative representations have
also been presented \cite{Alb05,Dig05}.  To determine the $U_{Q\bar
Q}^{(1)}$ potential as given by Eq.\ (\ref{Uqq1}), we also need the
ratio $a(T)=3p/\epsilon$ from the plasma equation of state.  We use
the quenched equation of state of Boyd $et~al.$ \cite{Boy96} for
quenched QCD, and the equation of state of Karsch $et~al.$
\cite{Kar00} for full QCD with 2 flavors. Using quark masses $m_c=1.4$
GeV and $m_b=4.3$ GeV, we can calculate the binding energies of
color-singlet heavy quarkonia and their spontaneous dissociation
temperatures.  We list in Table I the heavy quarkonium spontaneous
dissociation temperatures calculated with the $U_{Q\bar Q}^{(1)}$
potential, the $F_1$ potential, and the $U_1$ potential, in both
quenched QCD and full QCD.

The spontaneous dissociation temperatures of $J/\psi$ and $\chi_b$
obtained with the $U_{Q\bar Q}^{(1)}$ potential in quenched QCD are
found to be 1.62$T_c$ and $1.18T_c$ respectively.  On the other hand,
spectral analyses in quenched QCD give the spontaneous dissociation
temperature of 1.62-1.70$T_c$ for $J/\psi$ \cite{Asa03} and
1.15-1.54$T_c$ for $\chi_b$ \cite{Pet05}.  Thus, the $U_{Q\bar
Q}^{(1)}$ potential of Eq.\ (\ref{Uqq1}) gives spontaneous
dissociation temperatures that agree with those from spectral function
analyses.  This indicates that the $U_{Q\bar Q}^{(1)}$ potential,
defined as the linear combination of $U_1$ and $F_1$ in Eq.\
(\ref{Uqq1}), may be an appropriate $Q$-$\bar Q$ potential for
studying the stability of heavy quarkonia in quark-gluon plasma.

\begin{figure} [h]
\centering
\includegraphics[angle=0,scale=0.65]{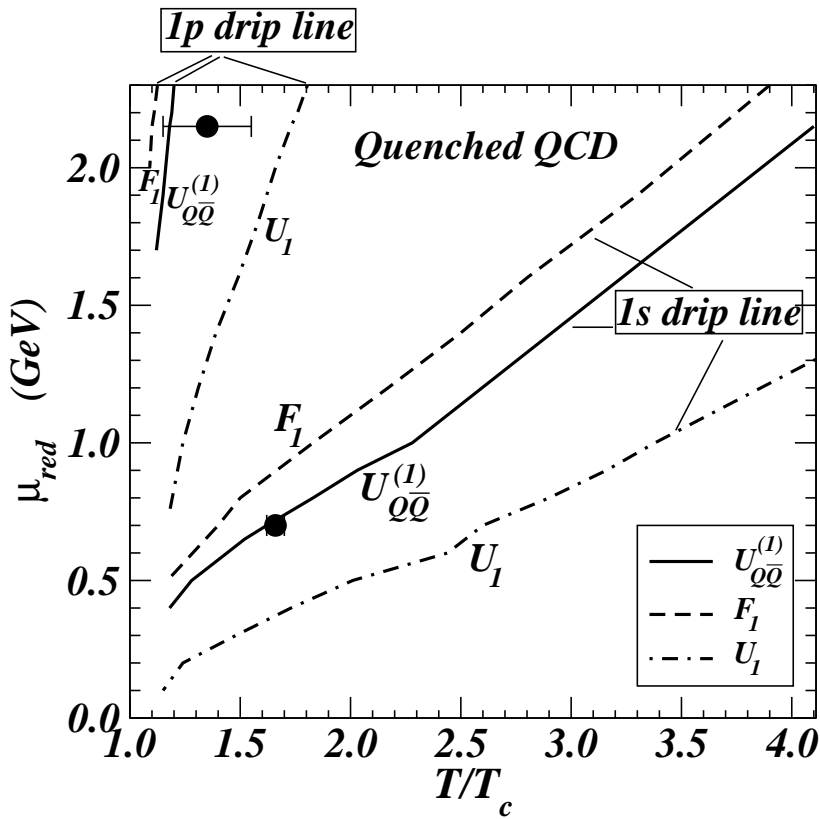}
\caption{The drip lines in quenched QCD calculated with the $U_{Q\bar
 Q}^{(1)}$, $F_1$, and $U_1$ potentials.  The solid-circle symbols
 represent results from lattice gauge spectral function analyses.}
\end{figure}

\section{The stability of a $Q$-$\bar Q$ pair}

To examine the stability of a $Q$-$\bar Q$ pair, we consider the quark
mass $m_Q$ as a variable and evaluate the spontaneous dissociation
temperature as a function of the reduced mass $\mu_{\rm red}=m_Q
m_{\bar Q}/(m_Q + m_{\bar Q})$. The results for quenched QCD are shown
in Fig. 1.  A state is bound in the $(T/T_c,\mu_{\rm red})$ space
above a drip line and is unbound below the drip line.  

The drip lines for quenched QCD in Fig.\ 1 are useful to provide a
comparison of the potential model results with those from spectral
analyses in the same quenched approximation.  However, as dynamical
quarks may provide additional screening, it is necessary to include
dynamical quarks to assess their effects on the stability of
quarkonia.  Accordingly, we use the $U_{Q\bar Q}^{(1)}$ potential
defined by Eq.\ (\ref{Uqq1}) with $F_1$, $U_1$, and $a(T)$ evaluated
in full QCD with 2 flavors \cite{Kac05,Kar00} to determine the drip
lines shown as solid curves in Fig.\ 2.  In comparison with quenched
QCD results, the $1s$ drip line in full QCD is shifted to lower
temperatures for $T>1.2T_c$ while the $1p$ drip line in full QCD is
only slightly modified.

\begin{figure} [h]
\centering
\includegraphics[angle=0,scale=0.65]{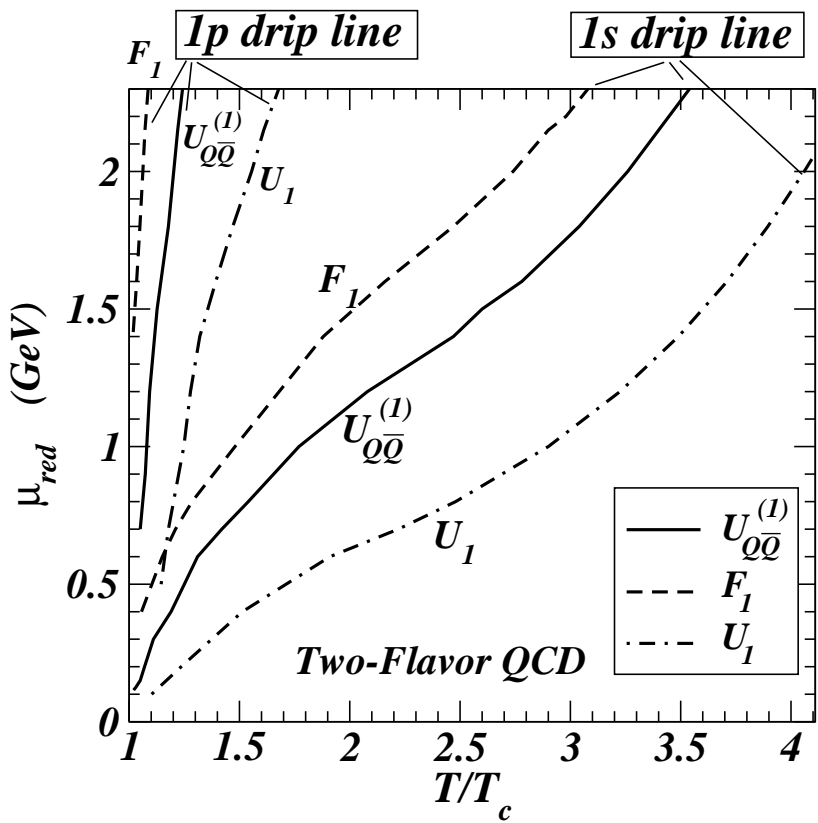}
\caption{The $1s$ and $1p$ quark drip lines calculated with the
$U_{Q\bar Q}^{(1)}$ potential in 2-flavor QCD (solid curves) and
quenched QCD (dashed curves).}
\end{figure}

For heavy quarkonia, the results in Table I and Fig. 2 indicate that
for the $U_{Q\bar Q}^{(1)}$ potential $J/\psi$, $\Upsilon$, and
$\chi_b$ may be bound in the plasma up to 1.42$T_c$, 3.40$T_c$, and
1.22$T_c$ respectively.  As light quarks have current-quark masses of
a few MeV, we may be advised that non-relativistic potential models
should not be used to describe light quarkonia. However, due to its
strong interaction with other constituents, a light quark becomes a
dressed quasiparticle and acquires a large quasiparticle mass.  In the
low temperature region when spontaneous chiral symmetry breaking
occurs with $\langle \bar \psi \psi\rangle\ne 0$, the quasiparticle
mass is $m_q\sim [|g\langle \bar \psi \psi\rangle|$+(current quark
mass)], where $g$ is the strong coupling constant and $\langle \bar
\psi \psi\rangle$ the quark condensate \cite{Hat85,Nam61,Szc01}.  This
quasiparticle mass is the origin of the constituent-quark mass in
non-relativistic constituent quark models \cite{Hat85,Szc01}.  In the
high temperature perturbative QCD region, the quasiparticle mass is
$m_q \sim gT/\sqrt{6}$, which is of the order of a few hundred MeV
\cite{Wel82}.

As the restoration of chiral symmetry is a second order transition,
$\langle \bar \psi \psi\rangle $ decreases gradually as the
temperature increases beyond $T_c$. The light quark quasiparticle mass
associated with $\langle \bar \psi \psi\rangle$ will likewise decrease
gradually from the constituent-quark mass to the current-quark mass
when the temperature increases beyond $T_c$.  This tendency for the
quasiparticle mass to decrease will be counterbalanced by the opposite
tendency for the quasiparticle `thermal mass' to increase with
increasing temperature.  In the region of our interest, $T_c<T<2T_c$,
various estimates give the light quark quasiparticle masses from 0.3
GeV to 1.2 GeV \cite{Lev98,Sza03,Iva05,Pet02}.  As in the case of
$T=0$, where light quarks with a constituent-quark mass of about 350
MeV mimic the effects of chiral symmetry breaking and non-relativistic
constituent quark models have been successfully used for light hadron
spectroscopy \cite{Szc01,Bar92}, so the large estimated quasiparticle
mass (from 0.3 to 1.2 GeV) may allow the use of a non-relativistic
potential model as an effective tool to estimate the stability of
light quarkonia at $T_c<T<2T_c$.  At this stage when the uncertainties
in the quasiparticle mass are much greater than the effects due to
relativistic kinematics, an estimate based on a non-relativistic model
suffices.  It will be of interest to investigate the relativistic
effects in the future, when the quasiparticle masses are more
definitively determined.

By examining the effects of the light quark quasiparticle masses on
the quark-gluon plasma equation of state, Levai $et~al.$ \cite{Lev98},
Szabo $et~al.$ \cite{Sza03}, and Ivanov $et~al.$ \cite{Iva05} estimate
that $m_q$ is about 0.3 to 0.4 GeV at $T_c<T<2T_c$.  From the results
in Fig. 2 for the $U_{Q\bar Q}^{(1)}$ potential, we can estimate that
as a quarkonium with light quarks has a reduced mass of 0.15-0.2 GeV,
it may be bound at temperatures below $(1.05-1.07)T_c$. An open heavy
quarkonium with a light quark and a heavy antiquark has a reduced mass
of about 0.3-0.4 GeV and may be bound at temperatures below
$(1.11-1.19)T_c$.

Another lattice gauge calculation gives $m_q/T=3.9\pm 0.2$ at $1.5T_c$
\cite{Pet02}, which implies that at $T$=1.5$T_c$ (or about 0.3 GeV),
the quark mass will be $\sim$1.2 GeV for $(u,d,s)$ quarks.  Such a
`light' quark quasiparticle mass appears to be quite large and may be
uncertain, as the plasma will have a relatively large abundance of
charm quarks and antiquarks, and there may be difficulties in
reproducing the plasma equation of state.  With this mass, a `light'
quarkonium will have a reduced mass of 0.6 GeV, and the quarkonium may
be bound at temperatures below $\sim$1.31$T_c$.

In either case, the drip lines of Fig. 2 calculated with the $U_{Q\bar
Q}^{(1)}$ potential for full QCD with 2 flavors do not support bound
$Q\bar Q$ states with light quarks beyond 1.5$T_c$. A recent study of
baryon-strangeness correlations suggests that the quark-gluon plasma
contains essentially no bound $Q\bar Q$ component at 1.5$T_c$
\cite{Koc05}.

In conclusion, we have used the thermodynamical quantities obtained in
lattice gauge calculations to determine the quark drip lines in
quark-gluon plasma for full QCD with two flavors.  The characteristics
of the quark drip lines severely limit the region of possible
quarkonium states with light quarks to temperatures close to the phase
transition temperature. If the light quark quasiparticle mass is
0.3-0.4 GeV as estimated from \cite{Lev98,Sza03,Iva05}, a quarkonium
with light quarks may be bound below (1.05-1.07)$T_c$ and an open
heavy quarkonium may be bound below (1.11-1.19)$T_c$.

\vspace*{0.3cm}
\vspace*{+0.3cm}
\noindent
{\bf Acknowledgment}

The author thanks Dr. H. Crater for helpful discussions.  This
research was supported in part by the Division of Nuclear Physics,
U.S. Department of Energy, under Contract No. DE-AC05-00OR22725,
managed by UT-Battelle, LLC and by the National Science Foundation
under contract NSF-Phy-0244786 at the University of Tennessee.

\end{document}